\documentclass[useAMS]{mn2e}

\usepackage{latexsym,graphicx}

\usepackage{color}

\newcommand\kms{{\rm\,km\,s^{-1}}}
\newcommand\msun{\rm\,M_\odot}

\newcommand\lsun{\rm\,L_\odot}

\newcommand\hii{H\,{\sc ii} \,}
\newcommand\myr{\msun \, {\rm yr}^{-1}}

\def\apgt{\ {\raise-.5ex\hbox{$\buildrel>\over\sim$}}\ }
\def\aplt{\ {\raise-.5ex\hbox{$\buildrel<\over\sim$}}\ }

\title[IRC\,$-$10\,414: a bow-shock-producing RSG star]{IRC\,$-$10414: a bow-shock-producing red supergiant
star\footnotemark[0]\thanks{Based on observation obtained with the
South African Large Telescope (SALT), programme
2011-3-RSA\_OTH-002.}}
\author[V.V.Gvaramadze et al.]
       {V. V.~Gvaramadze,$^{1,2}$\thanks{E-mail:
       vgvaram@mx.iki.rssi.ru}
       K. M.~Menten,$^{3}$ A. Y.~Kniazev,$^{4,5,1}$ N.~Langer,$^{6}$
       \newauthor J.~Mackey,$^{6}$ A.~Kraus,$^{3}$ D. M.-A.~Meyer$^{6}$ and T.~Kami\'{n}ski$^{3}$\\
        $^{1}$Sternberg Astronomical Institute, Lomonosov Moscow State University, Universitetskij Pr. 13, Moscow 119992, Russia\\
        $^{2}$Isaac Newton Institute of Chile, Moscow Branch, Universitetskij Pr. 13, Moscow 119992, Russia\\
        $^{3}$Max-Planck-Institut f\"ur Radioastronomie, Auf dem H\"ugel 69, 53121 Bonn, Germany\\
        $^{4}$South African Astronomical Observatory, PO Box 9, 7935 Observatory, Cape Town, South Africa \\
        $^{5}$Southern African Large Telescope Foundation, PO Box 9, 7935 Observatory, Cape Town, South Africa \\
        $^{6}$Argelander-Institut f\"ur Astronomie der Universit\"at Bonn, Auf dem H\"ugel 71, 53121, Bonn, Germany \\
        }
\begin{document}

\date{Accepted 2013 October 8.  Received 2013 September 25; in original form 2013 August 31}

\maketitle

\label{firstpage}

\begin{abstract}
Most runaway OB stars, like the majority of massive stars residing
in their parent clusters, go through the red supergiant (RSG)
phase during their lifetimes. Nonetheless, although many dozens of
massive runaways were found to be associated with bow shocks, only
two RSG bow-shock-producing stars, Betelgeuse and $\mu$\,Cep, are
known to date. In this paper, we report the discovery of an
arc-like nebula around the late M-type star IRC\,$-$10414 using
the SuperCOSMOS H-alpha Survey. Our spectroscopic follow-up of
IRC\,$-$10414 with the Southern African Large Telescope (SALT)
showed that it is a M7 supergiant, which supports previous claims
on the RSG nature of this star based on observations of its maser
emission. This was reinforced by our new radio- and
(sub)millimeter-wavelength molecular line observations made with
the Atacama Pathfinder Experiment (APEX) 12 meter telescope and
the Effelsberg 100 m radio telescope, respectively. The SALT
spectrum of the nebula indicates that its emission is the result
of shock excitation. This finding along with the arc-like shape of
the nebula and an estimate of the space velocity of IRC\,$-$10414
($\approx70\pm20 \, \kms$) imply the bow shock interpretation for
the nebula. Thus, IRC\,$-$10414 represents the third case of a
bow-shock-producing RSG and the first one with a bow shock visible
at optical wavelengths. We discuss the smooth appearance of the
bow shocks around IRC\,$-$10414 and Betelgeuse and propose that
one of the necessary conditions for stability of bow shocks
generated by RSGs is the ionization of the stellar wind. Possible
ionization sources of the wind of IRC\,$-$10414 are proposed and
discussed.
\end{abstract}

\begin{keywords}
Stars: kinematics and dynamics -- stars: massive -- circumstellar
matter -- stars: individual: (IRC\,$-$10414, WR\,114, Betelgeuse).
\end{keywords}

\section{Introduction}
\label{sec:intro}

Massive stars form in compact star clusters (Lada \& Lada 2003)
and then find themselves in the field due to three major
processes: few-body dynamical encounters (Poveda, Ruiz \& Allen
1967; Gies \& Bolton 1986), binary-supernova explosions (Blaauw
1961; Stone 1991) and cluster dissolution (Tutukov 1978; Kroupa,
Aarseth \& Hurley 2001). Unlike the last process, the first two
ones can produce stars with very high (from tens to hundreds
$\kms$) space velocities (Leonard 1991; Portegies Zwart 2000;
Gvaramadze 2009; Gvaramadze, Gualandris \& Portegies Zwart 2009).
Some of these, so-called runaway (Blaauw 1961), stars move
supersonically through the interstellar medium (ISM) and give rise
to bow shocks ahead of them. The bow shocks have a characteristic
arc-like shape and can be detected in the optical (Gull \& Sofia
1979), infrared (van Buren \& McCray 1988), radio (Benaglia et al.
2010), and X-ray (L\'{o}pez-Santiago et al. 2012) wavebands.

Interestingly, the vast majority of bow-shock-producing stars are
either on the main-sequence or are blue supergiants (BSGs), while
there are no Wolf-Rayet (WR) stars associated with distinct bow
shocks and only two known bow-shock-producing red supergiants
(RSGs), Betelgeuse (Noriega-Crespo et al. 1997b) and $\mu$\,Cep
(Cox et al. 2012). Moreover, most of these bow shocks were
discovered in the mid-infrared and there are only few unambiguous
detections in optical (Gull \& Sofia 1979; Hollis et al. 1992;
Kaper et al. 1997; Gvaramadze \& Bomans 2008; Gvaramadze et al.
2011b) and other wave-bands (Benaglia et al. 2010;
L\'{o}pez-Santiago et al. 2012). Detection of optical bow shocks
and their follow-up spectroscopy can be used for constraining
parameters of the stellar wind and the local interstellar
environment (e.g. Kaper et al. 1997; Hollis et al. 1992), and for
estimating the shock and thereby the stellar space velocities,
which in turn might serve as a basis for confronting results of
numerical modelling of stellar wind bow shocks with observations
(Meyer et al., in preparation).

In this paper, we report the discovery of an optical arc-like
nebula around the late M-type star IRC\,$-$10414 and propose a
possible explanation for its origin. We review the existing data
on IRC\,$-$10414 in Section\,\ref{sec:irc} and present the
arc-like nebula in Section\,\ref{sec:arc}. Spectroscopic
observations of IRC\,$-$10414 and the nebula, and the results of
these observations are described and discussed in
Section\,\ref{sec:spec}. In Section\,\ref{sec:maser} we present
and discuss our observation of the maser emission from
IRC\,$-$10414. IRC\,$-$10414 and its nebula are further discussed
in Section\,\ref{sec:dis}. We summarize in Section\,\ref{sec:sum}.

\section{IRC\,$-$10414: observational data}
\label{sec:irc}

IRC\,$-$10414 (also known as IRAS\,18204$-$1344, RAFGL\,2139 and
OH\,17.55$-$0.13) is a late M-type star and a prominent source of
OH (Kolena \& Pataki 1977), H$_2$O (Kleinmann, Sargent \&
Dickinson 1978; Lada et al. 1981) and SiO maser emission (Ukita \&
Goldsmith 1984). Hansen \& Blanco (1975) classified IRC\,$-$10414
as an M8 star using objective-prism spectroscopy. Lockwood (1985)
derived for IRC\,$-$10414 a spectral type of M6.5\,III from visual
and near-infrared photometry, while Kwok, Volk \& Bidelman (1997)
inferred a type of M7 on the basis of the {\it InfraRed
Astronomical Satelite} ({\it IRAS}) low-resolution spectrum.

Assuming that IRC\,$-$10414 is an asymptotic giant branch (AGB)
star with a luminosity of $L=10^4 \, \lsun$, Jura \& Kleinmann
(1989) derived a distance to the star of $d=0.7$ kpc. This
distance estimate is often quoted in the literature. Jura \&
Kleinmann (1989) also derived the mass-loss rate of IRC\,$-$10414
of $\dot{M}\approx 4\times 10^{-6} \, \myr$ using the prescription
by Jura (1987), which is based on a scaling of the {\it IRAS}
60\,$\mu$m flux density.

Ukita \& Goldsmith (1984) were the first to suggest that
IRC\,$-$10414 is a RSG. Their suggestion was based on the large
width of a single SiO maser emission profile and large circular
polarization and velocity width of the OH maser emission, which
are more typical of RSG than AGB stars (e.g. Le Bertre \& Nyman
1990; Verheyen, Messineo \& Menten 2012; see also
Section\,\ref{sec:mas-res}).

Ukita \& Goldsmith (1984) also discussed a possible membership of
IRC\,$-$10414 of the OB association Sct\,OB3, but rejected it
because the kinematic distance to IRC\,$-$10414 of $\approx 4.5$
kpc, derived from the local-standard of rest (LSR) velocity of the
star of $v_{\rm LSR}=42 \, \kms$ (cf. Engels 1979; Lada et al.
1981; see also Section\,\ref{sec:mas-res}), is significantly
larger than the distance to Sct\,OB3 of $\approx 1.7$ kpc
(Humphreys 1978). Using the kinematic distance they re-estimated
the luminosity of IRC\,$-$10414 to be $L\approx 4\times 10^5 \,
\lsun$, thereby further supporting the RSG nature of this star.

\begin{table}
  \caption{Details of IRC\,$-$10414.}
  \label{tab:det}
  \begin{center}
 \begin{tabular}{lrr}
  \hline
  Spectral type & M7\,I \\
  RA(J2000) & $18^{\rm h} 23^{\rm m} 17\fs90$ \\
  Dec.(J2000) & $-13\degr 42\arcmin 47\farcs3$ \\
  $l, b$ & $17\fdg5509$, $-0\fdg1262$ \\
  $V$ (mag) & $\approx 12$ \\
  $J$ (mag) & 2.845$\pm$0.290 \\
  $H$ (mag) & 1.403$\pm$0.284 \\
  $K_{\rm s}$ (mag) & 0.713$\pm$0.336 \\
  $v_{\rm LSR}$ ($\kms$) & $43\pm2$ \\
  $v_\infty$ ($\kms$) & $21\pm2$ \\
  $P$ (d) & 768.16; 2726.43 \\
  \hline
 \end{tabular}
\end{center}
\end{table}
\begin{figure*}
\includegraphics[width=17.5cm]{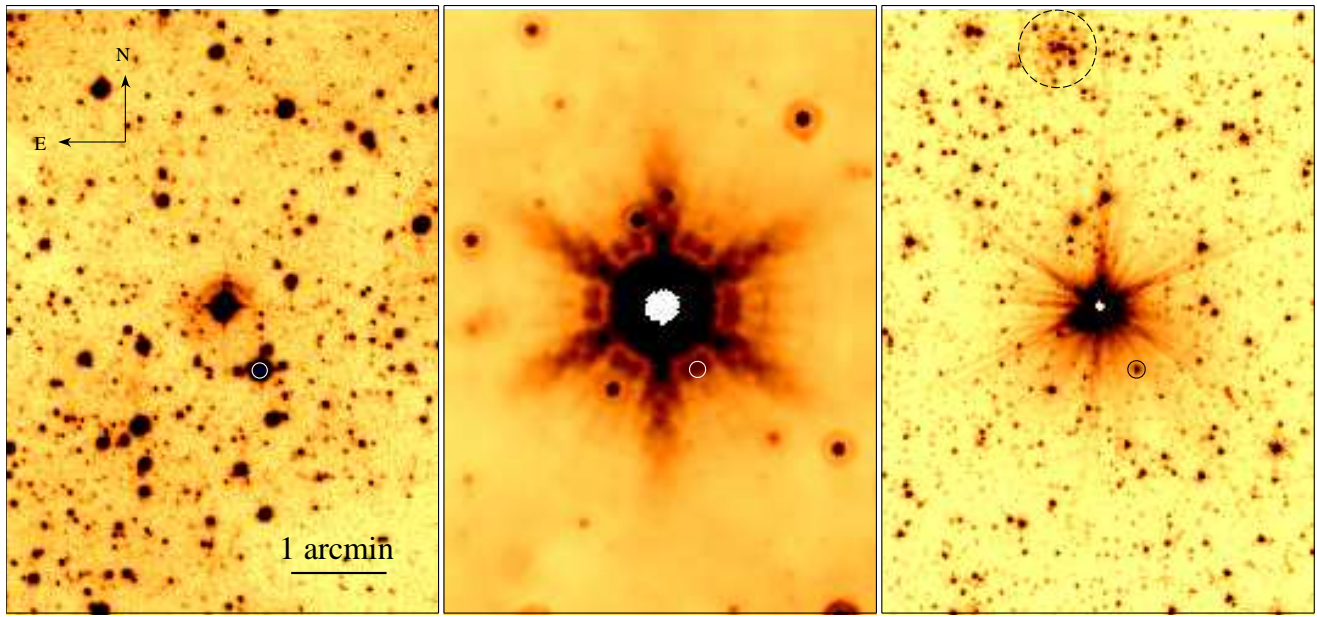}
\centering \caption{From left to right: SHS and {\it Spitzer} MIPS
24\,$\mu$m and IRAC 3.6\,$\mu$m images of the region around
IRC\,$-$10414 and its associated arc-like nebula (both {\it
Spitzer} images are highly saturated). The orientation and the
scale of the images are the same. The position of the WC5 star
WR\,114 (located at $\approx$45 arcsec southwest of IRC\,$-$10414)
is marked by a circle. A stellar concentration visible in the
3.6\,$\mu$m image at $\approx$2 arcmin to the north of
IRC\,$-$10414 (marked by a dashed circle) is the background
globular cluster Mercer\,5. See text for details. At a distance of
2 kpc, 1 arcmin corresponds to $\approx$0.57 pc.} \label{fig:arc}
\end{figure*}

The RSG interpretation of IRC\,$-$10414 has recently been
reinforced by Maeda et al. (2008). Using five-epoch very long
baseline interferometry (VLBI) observations of IRC\,$-$10414,
these authors measured proper motions of knots of the H$_2$O maser
emission and estimated the distance to the star to be $d\approx2$
kpc. Correspondingly, the luminosity and the mass-loss rate of
IRC\,$-$10414 given in Jura \& Kleinmann (1989) were scaled up to
much larger values of $L\approx 10^5 \, \lsun$ and $\dot{M}\approx
3\times 10^{-5} \, \myr$, respectively.

In the following, we adopt a distance to IRC\,$-$10414 of $d=2$
kpc. This distance is much smaller than the kinematic one derived
by Ukita \& Goldsmith (1984). The discrepancy could be resolved if
IRC\,$-$10414 is a runaway star, i.e. if it has a significant
peculiar radial velocity in addition to the radial velocity purely
determined by circular rotation of the Galaxy, which is a natural
assumption given the location of this star in the field (cf.
Gvaramadze et al. 2012). This would essentially prohibit any
``kinematic'' distance determination\footnote{See, e.g., the
famous (factor of 2) case of the kinematic distance of the giant
molecular cloud W3 vs. the spectro-photometric distance of the
Per\,OB associations (Xu et al. 2006).}. The runaway status of
IRC\,$-$10414 is supported by the detection of a bow shock around
this star (see Sections\,\ref{sec:arc} and \ref{sec:spec-bow} for
details) and by an estimate of its space velocity
(Section\,\ref{sec:run}).

IRC\,$-$10414 is indicated in the International Variable Star
Index (VSX) as a variable (Watson, Henden \& Price 2006) with the
median $V$-band magnitude of 11.96 and the period of variation and
its amplitude of $P=768.16$ d and $\Delta V=1.32$ mag,
respectively. These figures were derived by Pojma\'{n}ski \&
Maciejewski (2005) from the light curve of IRC\,$-$10414 obtained
from the All-Sky Automated Survey (ASAS). The period and amplitude
of pulsations were recently revised by Richards et al. (2012), who
used the ASAS $V$-band light curve on a much longer time interval
to derive $P\approx 2726.43$ d and $\Delta V=1.19$ mag. These
authors also provided probabilistic classification of
IRC\,$-$10414 with probability of being a semi-regular pulsating
variable, RSG or Mira of 0.3450, 0.2744 and 0.1280, respectively.

Using the peak separation of 1612 MHz masing OH lines, Blommaert,
van Langevelde \& Michiels (1994) and Sevenster et al. (2001)
estimated the terminal velocity of the stellar wind of
IRC\,$-$10414 to be $v_\infty\approx 15-16 \, \kms$ (see also
Section\,\ref{sec:mas-res}).

The details of IRC\,$-$10414 are summarized in
Table\,\ref{tab:det}. The spectral type is from
Section\,\ref{sec:spec-irc}. The coordinates and the $J$, $H$,
$K_{\rm s}$ magnitudes are taken from the 2MASS (Two Micron All
Sky Survey) All-Sky Catalog of Point Sources (Cutri et al. 2003).
$v_{\rm LSR}$ and $v_\infty$ are from Section\,\ref{sec:mas-res}.

\section{Arc-like nebula around IRC\,$-$10414}
\label{sec:arc}

The arc-like nebula around IRC\,$-$10414 was serendipitously
discovered during our search for bow shocks generated by OB stars
running away from the young massive star cluster NGC\,6611 (for
motivation and the results of this search, see Gvaramadze \&
Bomans 2008). Using the SuperCOSMOS H-alpha Survey\footnote{This
survey provides narrow-band H$\alpha$ plus [N\,{\sc ii}]
$\lambda\lambda$6548, 6584 images of the Southern Galactic Plane
and Magellanic Clouds with resolution of $\sim$1 arcsec.} (SHS;
Parker et al. 2005), we searched for possible optical counterparts
to bow shocks detected in the archival data of the Mid-Infrared
Galactic Plane Survey (performed by the {\it Midcourse Space
Experiment} satellite; Price et al. 2001) and found an arc-like
structure (see the left panel of Fig.\,\ref{fig:arc}) at
$\approx$1.1 degree (or $\approx 38$ pc in projection) to the east
of the cluster (see Section\,\ref{sec:run}). We interpret this arc
as a bow shock, as further discussed in
Section\,\ref{sec:spec-bow}. The orientation of the bow shock,
however, is inconsistent with the possibility that its associated
star was ejected from NGC\,6611 (see Section\,\ref{sec:run}).
Using the SIMBAD data
base\footnote{http://simbad.u-strasbg.fr/simbad/}, we identified
this star with the late M-type one IRC\,$-$10414. We found also
that IRC\,$-$10414 is located at only $\approx$45 arcsec from the
WC5 (Smith 1968) star WR\,114 (see Fig.\,\ref{fig:arc}). A
possible physical relationship between the two stars is discussed
in Sections\,\ref{sec:run} and \ref{sec:ion}.

The stand-off distance of the bow shock, i.e. the minimum distance
from the star at which the ram pressure of the stellar wind is
balanced by the ram pressure of the ambient ISM, is $R_{\rm
SO}=\theta d$, where $\theta\approx$15 arcsec is the angular
separation between the apex of the bow shock and the star. At
$d=2$ kpc, $R_{\rm SO}\approx$0.14 pc; we neglected here a small
geometrical factor ($\sim1$) taking into account inclination of
the bow shock to the plane of the sky (see
Section\,\ref{sec:run}).

The bow shock interpretation of the arc implies that it should
also be visible in the mid-infrared (e.g. van Buren \& McCray
1988; Noriega-Crespo, van Buren \& Dgani 1997a; Peri et al. 2012),
while its relatively large angular extent suggests that it could
be resolved with the Multiband Imaging Photometer for {\it
Spitzer} (MIPS; Rieke et al. 2004) and the {\it Wide-field
Infrared Survey Explorer} ({\it WISE}; Wright et al. 2010), which
provide 24 and 22\,$\mu$m images at 6 and 12 arcsec resolution,
respectively (cf. Gvaramadze, Kroupa \& Pflamm-Altenburg 2010;
Gvaramadze, Pflamm-Altenburg \& Kroupa 2011a). Unfortunately, the
very high infrared brightness of IRC\,$-$10414 made detection of
an infrared counterpart to the arc impossible because of heavy
saturation of the MIPS and {\it WISE} images. This is illustrated
in the middle panel of Fig.\,\ref{fig:arc}, where we present the
MIPS 24\,$\mu$m image of the star obtained within the framework of
the 24 and 70 Micron Survey of the Inner Galactic Disk with MIPS
(MIPSGAL; Carey et al. 2009).

IRC\,$-$10414 is also heavily saturated in all four (3.6, 4.5,
5.8, and 8 $\mu$m) images obtained with the Infrared Array Camera
(IRAC; Fazio et al. 2004) within the Galactic Legacy Infrared
Mid-Plane Survey Extraordinaire (GLIMPSE; Benjamin et al. 2003).
In the 3.6\,$\mu$m image we found a stellar concentration at
$\approx$2.8 arcmin to the north of IRC\,$-$10414 (see the right
panel of Fig.\,\ref{fig:arc}), which turns out to be the
background globular cluster Mercer\,5 (Mercer et al. 2005;
Longmore et al. 2011).

IRC\,$-$10414 was also covered by the Multi-Array Galactic Plane
Imaging Survey (MAGPIS; Helfand et al. 2006) carried out with the
Very Large Array (VLA), and by the {\it X-ray Multi-Mirror
Mission} ({\it XMM-Newton}) observation of WR\,114 (Oskinova et
al. 2003). No radio or X-ray counterpart to the arc and the star
were detected. On the other hand, we found that IRC\,$-$10414 is
located at $\approx$1 arcmin to the northeast of the $\sim$6
arcmin diameter (radio) shell of the supernova remnant (SNR)
G017.4-00.1 (Brogan et al. 2006), which most probably is a
background object (Pavlovi\'{c} et al. 2013).

\section{Spectroscopy of IRC\,$-$10414 and its associated
arc-like nebula}
\label{sec:spec}

\subsection{Spectroscopic observations and data reduction}
\label{sec:spec-obs}

\begin{figure}
\includegraphics[width=6cm,angle=270]{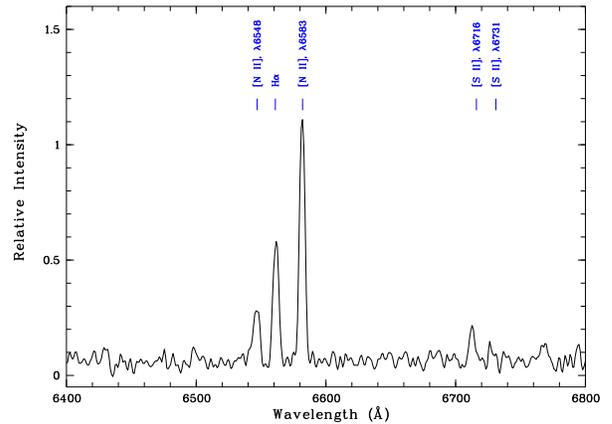}
\caption{Observed spectrum of the arc-like nebula associated with
IRC\,$-$10414 in the region of the H$\alpha$ line.}
\label{fig:spec-2}
\end{figure}

IRC\,$-$10414 and the arc-like nebula were observed within the
framework of our ongoing programme of spectroscopic follow-up of
bow-shock-producing stars (Gvaramadze et al. 2011b, 2013). The
observation was conducted with the Southern African Large
Telescope (SALT; Buckley, Swart \& Meiring 2006; O'Donoghue et al.
2006) on 2012 April 20 with the Robert Stobie Spectrograph (RSS;
Burgh et al. 2003; Kobulnicky et al. 2003). The long-slit
spectroscopy mode of the RSS was used with a 1.25 arcsec slit
width. The slit was oriented along the line connecting
IRC\,$-$10414 and WR\,114 (position angle $\approx30$ degrees; see
Fig.\,\ref{fig:arc}). The grating GR900 was used to cover a total
spectral range of 4200$-$9100 \AA \, with a final reciprocal
dispersion of $0.97$ \AA \, pixel$^{-1}$ and FWHM spectral
resolution of 4.74$\pm$0.05 \AA. The seeing during the observation
was $\approx$1.5 arcsec. IRC\,$-$10414 was observed with two 600 s
exposures for the blue (4200$-$7250 \AA) part of the spectrum and
one 60 s exposure for the red (6420$-$9100 \AA) one. The spectrum
of a Xe comparison arc was obtained to calibrate the wavelength
scale as well as spectral flats to correct for pixel-to-pixel
variations. A spectrophotometric standard star was observed after
observation of the object for relative flux calibration. Primary
reduction of the data was done with the SALT science pipeline
(Crawford et al. 2010). After that, the bias and gain corrected
and mosaicked long-slit data were reduced in the way described in
Kniazev et al. (2008). The red end of the spectrum ($>8400$ \AA)
shows significant fringing effects, which were corrected in the
way described in Kniazev et al. (2013). The final, totally reduced
and corrected for sensitivity, spectra of IRC\,$-$10414 and the
arc-like nebula are shown in Figs.\,\ref{fig:spec} and
\ref{fig:spec-2}.

\subsection{Spectral classification of IRC\,$-$10414}
\label{sec:spec-irc}

\begin{figure}
\includegraphics[width=6cm,angle=270]{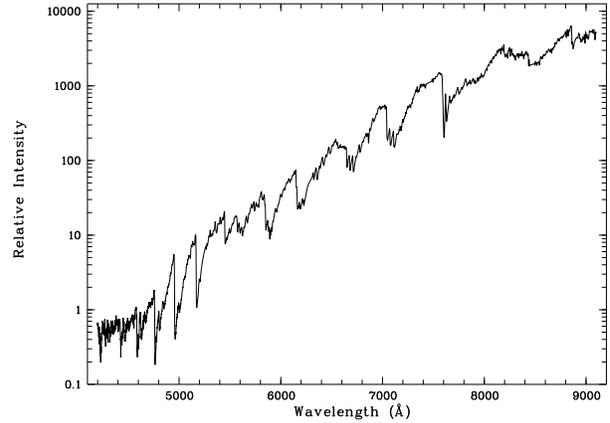}
\caption{Observed spectrum of IRC\,$-$10414.}
\label{fig:spec}
\end{figure}

Fig.\,\ref{fig:spec} shows that the spectrum of IRC\,$-$10414 is
dominated by very strong molecular TiO absorption bands, which are
typical of late M-type stars. The presence of the weak VO bands at
$\lambda\lambda$5737, 6478, 7400, 7961 and 7900 implies that
IRC\,$-$10414 is an M7 star (Solf 1978; Turnshek et al. 1985),
which agrees with the classification of this star based on the
low-resolution {\it IRAS} spectrum (Kwok et al. 1997).

There are no solid criteria for assigning luminosity classes to
late M stars because of paucity and variability of these stars. To
determine the luminosity class for IRC\,$-$10414, we rely on the
luminosity sensitive Ca\,{\sc ii} $\lambda\lambda$8498, 8542, 8662
triplet, which is known to be a good luminosity indicator for
early M stars (e.g. Jaschek \& Jaschek 1987). It was found that
the sum of EWs of lines of this triplet shows a good correlation
with the stellar gravity, $g$, for stars with spectral type from F
to M5 (Jones, Alloin \& Jones 1984). Extrapolation of this
correlation might be useful for constraining the luminosity
classes of late M stars. Measurement of EWs of the Ca\,{\sc ii}
triplet lines in the late M stars however is challenging because
the TiO bands significantly affect their spectra, resulting in the
depression of the continuum level and leading to the apparent
reduction of EWs of the Ca\,{\sc ii} lines (e.g. Sharpless 1956;
Zhu et al. 1999).

To measure EWs of the Ca\,{\sc ii} triplet, we follow the approach
by Jones et al. (1984) and approximate the continuum between 8430
and 8852 \AA \, as a straight line (cf. Beauchamp, Moffat \&
Drissen 1994). In doing so, we derived EW(8498)$\approx5.5$ \AA,
EW(8542)$\approx6.5$ \AA \, and EW(8662)$\approx5.5$ \AA \, and
estimated an uncertainty of $\approx2$ \AA \, for these values.
The total EW of the three Ca\,{\sc ii} lines of $17.5\pm3.5$ and
fig.\,3 in Jones et al. (1984) suggest that $\log g$ of
IRC\,$-$10414 should be negative (see also
Section\,\ref{sec:par}), which in turn implies the luminosity
class of I for this star (cf. Beauchamp et al. 1994). Thus,
IRC\,$-$10414 is an M7\,I star.

\subsection{Arc-like nebula as a bow shock}
\label{sec:spec-bow}

\begin{table}
\centering{ \caption{Relative intensities and RVs of lines in the
spectrum of the arc-like nebula associated with IRC\,$-$10414.}
\label{tab:int}
\begin{tabular}{lcc} \hline
\rule{0pt}{10pt} $\lambda$(\AA) Ion  & I($\lambda$)/I(H$\alpha$) &
RV ($\kms$) \\
\hline
6548\ [N\ {\sc ii}]\  & 0.53$\pm$0.05  &  $-$22.1$\pm$8.1 \\
6563\ H$\alpha$\      & 1.00$\pm$0.04  &  $-$1.4$\pm$2.9  \\
6583\ [N\ {\sc ii}]\  & 1.89$\pm$0.04  &  $-$16.7$\pm$1.8 \\
6716\ [S\ {\sc ii}]\  & 0.28$\pm$0.03  &  $-$10.9$\pm$9.9 \\
6731\ [S\ {\sc ii}]\  & 0.12$\pm$0.03  &  $-$9.1$\pm$19.5 \\
\hline
\end{tabular}
 }
\end{table}

Fig.\,\ref{fig:spec-2} presents the spectrum of the arc-like
nebula in the same scale as the spectrum of IRC\,$-$10414 in
Fig.\,\ref{fig:spec}. The comparison of the two spectra shows that
in the region of the H$\alpha$ line the nebula is more than 6 mag
fainter than the star. This, along with the proximity of the
nebula to the star and the high extinction towards them ($\approx
5$ mag in the visual range; see Section\,\ref{sec:par}), makes the
spectroscopy of the former object a non-trivial observational
task, which is impracticable with small telescopes. Even with the
SALT we were not able to detect any emission lines in the blue
spectrum of the arc. In the red part of the spectrum, we found the
H$\alpha$ emission line and two forbidden emission doublets of
[N\,{\sc ii}] $\lambda\lambda$6548, 6583 and [S\,{\sc ii}]
$\lambda\lambda$6716, 6731, of which the nitrogen doublet is
particularly strong (see Fig.\,\ref{fig:spec-2}). Relative
intensities and heliocentric radial velocities (RVs) of these
lines (measured by fitting Gaussians to them) are given in
Table\,\ref{tab:int}.

With so few detected lines in the spectrum of the arc-like nebula
it is not possible to derive quantitative estimates of its
elemental abundances, the electron temperature, $T_{\rm e}$, or
the extinction towards the arc. Still, some useful information can
be extracted from the spectrum.

The ratio of the combined [S\,{\sc ii}] $\lambda\lambda$6716, 6731
lines against H$\alpha$ can be used for distinguishing
shock-heated nebulae (e.g. SNRs and Herbig-Haro objects) from
photoionized ones [e.g. \hii regions and most planetary nebulae
(PNe)]. In SNRs this ratio is enhanced with respect to that
measured for \hii regions and typically is $\geq 0.3-0.4$ (e.g.
Fesen, Blair \& Kirshner 1985; Stupar \& Parker 2009; Leonidaki,
Boumis \& Zezas 2013). For the arc-like nebula we found the
[S\,{\sc ii}]/H$\alpha$ ratio of 0.4, which suggests that the
ionization of the nebula is because of shock-heating.

The high intensity ratio of the [S\,{\sc ii}] doublet relative to
H$\alpha$ is also observed in old, low-density PNe, which show
signatures of interaction with the ISM (e.g., Sh\,2-176: Sabbadin,
Minello \& Bianchini 1977; Sh\,2-216: Fesen, Blair \& Gull 1981;
Reynolds 1985; PFP\,1: Pierce et al. 2004) and whose emission can
also be attributed to shock-heating. The PN nature of the arc,
however, can be rejected on the basis of the plot of the [S\,{\sc
ii}] electron number density (as measured from the [S\,{\sc ii}]
$\lambda\lambda$6716, 6731 line intensity ratio; see below),
$n_{\rm e}$, versus the radius for PNe given in Phillips (1998;
see fig.\,5 therein). According to this plot, a PN with $n_{\rm e}
\la 100 \, {\rm cm}^{-3}$  must have a radius of $\ga 0.3$ pc.
This would indicate a distance to IRC\,$-$10414 of $\ga 4$ kpc and
a luminosity of this star of $\ga 6\times10^5 \, {\rm L}_{\odot}$,
which is too high for central stars of PNe. Since the arc around
IRC\,$-$10414 is definitely not a SNR, nor a Herbig-Haro object,
we conclude that it is a bow shock. The bow shock interpretation
of the arc-like nebula is also supported by the runaway status of
IRC\,$-$10414 (see Section\,\ref{sec:run}), and by morphology of
the nebula itself.

The [S\,{\sc ii}] $\lambda\lambda$6716, 6731 line intensity ratio
can be used to derive $n_{\rm e}$ in the line-emitting region
(Krueger, Aller \& Czyzak 1970; Saraph \& Seaton 1970). Note that
the measured ratio of 2.33$\pm$0.63 significantly exceeds the
theoretical one of $\approx 1.4$ (Krueger et al. 1970) and the one
based on the apparent peak intensities (see
Fig.\,\ref{fig:spec-2}). This discrepancy is due to the weakness
of the [S\,{\sc ii}] $\lambda$6731 line, whose shape is strongly
affected by the noise. Nevertheless, the measured and theoretical
ratios are in agreement with each other at 1.4 sigma level, and
both are in the low-density limit. Thus, we can only put an upper
limit on $n_{\rm e}$ of $\leq 100 \, {\rm cm}^{-3}$.

The [N\,{\sc ii}] and [S\,{\sc ii}] line intensities can be used
to estimate the nitrogen to sulphur abundance ratio, which is
almost independent of $n_{\rm e}$ and $T_{\rm e}$, provided that
$n_{\rm e}\leq 100 \, {\rm cm}^{-3}$ and $T_{\rm e}\leq 10^4$ K.
In this case, the abundance ratio is given by (see Benvenuti,
D'Odorico \& Peimbert 1973 and references therein):
\begin{equation}
{N({\rm N}^+ )\over N({\rm S}^+ )} =3.61 {I(6584) \over
I(6716+6731)} \, . \label{eq:abud}
\end{equation}
Unfortunately, the spectrum of the arc is not deep enough to
detect the electron temperature diagnostic line [N\,{\sc ii}]
$\lambda$5755. Nonetheless, it is likely that $T_{\rm e}$ in the
line-emitting region is $\leq 10^4$ K. Using
equation\,(\ref{eq:abud}) and Table\,\ref{tab:int}, one finds
$N({\rm N}^+ )/N({\rm S}^+ )\approx 17$, which more than three
times exceeds the solar value of 5.12 (Asplund et al. 2009). This
estimate suggests that the line-emitting material is enriched in
nitrogen and that its composition might reflect that of the
stellar wind (cf. Reimers et al. 2008), which in RSGs is enhanced
in N by a factor of $\sim 3-5$ because of the dredge-up of the
nuclear processed gas (e.g. Brott et al. 2011). We conclude
therefore that the emission of the bow shock at least partially
originates in the shocked stellar wind.

The overabundance of N could be responsible for the very high
[N\,{\sc ii}]/H$\alpha$ ratio observed in the arc. Such high line
ratios were also detected in circumstellar nebulae around some
evolved massive stars, e.g. the WN8h star WR\,124 (Esteban et al.
1991) and the luminous blue variable (LBV) star AG\,Car (Smith et
al. 1997), which might be considered as the evidence that these
stars have evolved through the red supergiant phase (e.g. Smith et
al. 1997; but see Lamers et al. 2001 for a different explanation
of enhanced N abundance in LBV nebulae). Similarly, the high
[N\,{\sc ii}]/H$\alpha$ ratios observed in some SNRs, e.g. W50
(Kirshner \& Chevalier 1980) and SNR\,G279.0+1.1 (Stupar \& Parker
2009), might be due to supernova explosions within pre-existing
wind-driven shells (created by supernova progenitor stars), which
are enriched by the RSG wind material.

Note that the high [N\,{\sc ii}]/H$\alpha$ ratio might not be
solely caused by the enhanced nitrogen abundance. It can also be
inherent to spectra of photoionized nebulae of normal chemical
composition, e.g., ionized regions surrounding the so-called
supersoft X-ray sources (Rappaport et al. 1994) or ionization
fronts (Henney et al. 2005). Note also that according to models
presented in the two afore-cited papers, the high-intensity
nitrogen lines in the spectra of photoionized nebulae should be
accompanied by very strong [O\,{\sc i}] $\lambda$6300 emission
line (stronger than or comparable to the H$\alpha$ line). However,
we did not detect this line in the spectrum of the nebula around
IRC\,$-$10414, which implies that its intensity is less than
$10-20$ per cent of that of the H$\alpha$ line.

To summarize, the sum of evidence strongly supports the view that
the arc around IRC\,$-$10414 is a bow shock.

\section{Molecular line emission from IRC\,$-$10414}
\label{sec:maser}

\subsection{Millimeter-wavelength and radio observations}
\label{sec:mas-obs}

\begin{figure}
\includegraphics[width=7.5cm]{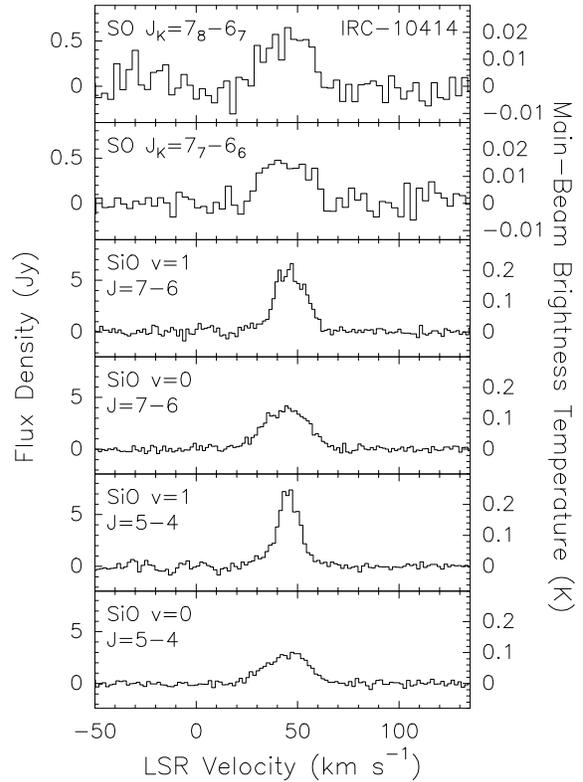}
\centering \caption{Sulfur monoxide (SO) and Silicon monoxide
(SiO) lines observed toward IRC\,$-$10414 with APEX. Bottom to
top, spectra of the $v=0$ and 1, $J=5-4$ and $v=0$ and 1, $J=7-6$
transition are plotted. The left and right abscissa show a flux
density and a main-beam-brightness temperature scale,
respectively. For clarity reasons, the SO lines were smoothed to a
channel spacing of $\approx 3.0$~km~s$^{-1}$ and the SiO lines to
$\approx 1.5$~km~s$^{-1}$. } \label{fig:sio}
\end{figure}

The millimeter-wavelength observations were made with the Atacama
Pathfinder Experiment (APEX) 12 meter telescope (G\" {u}sten et
al. 2006) under (for its Atacama desert site) moderate weather
conditions. On 2012 April 4 and May 8 we used the APEX-1 receiver,
which is part of the Swedish heterodyne facility instrumentation
(Vassilev et al. 2008), to observe the SiO $J=5-4$ rotational
transition, both in the vibrational ground and the first excited
state ($v=0$ and 1);  see Table\,\ref{tab:apex} for information on
the observed lines. The data were calibrated using the standard
chopper wheel method. Employing the 2.5 GHz total bandwidth of the
newest version of the APEX facility Fast Fourier Transform
Spectrometer (FFTS; Klein et al. 2006) available at the time, each
($v=0$ and 1) pair of lines could be observed in one band pass.
The FFTS's 2.5 GHz were split into 32768 channels with a spacing
of 76.3 KHz, which corresponds to 0.105 and $0.088 \, \kms$ at 217
and 260 GHz, respectively. For clarity reasons the spectra shown
in Fig.\,\ref{fig:sio} have been to binned in $\approx 1.6 \,
\kms$ wide channels. On 2013 August 14, we used an upgraded
version of the First Light APEX Submillimeter Heterodyne
instrument (FLASH; Heyminck et al. 2006), which is an MPIfR
principal investigator instrument, to observe the SiO $v=0$ and 1
$J=7-6$ lines around 302 GHz. Like APEX-1, FLASH is a two sideband
(2SB) receiver. However, its bandwidth is a larger 4 GHz for each
sideband. The wide bandwidth allowed serendipitous detection of
two lines from sulfur monoxide (SO); see Table\,\ref{tab:apex}.

Conversion to flux density units (in Jy) and a main-beam
brightness temperature scale (in K) was established by
interpolating the aperture and main-beam efficiencies, $\eta_{\rm
A}$ and $\eta_{MB}$, given by Vassilev et al. (2008) and G{\"
u}sten et al. (2006). Values are $\eta_{\rm A}=0.67$ and
$\eta_{\rm MB}=0.82$ for the SiO $J=5-4$ lines, and $\eta_{\rm
A}=0.60$ and $\eta_{\rm MB}=0.73$ for the $J=7-6$ and the SO
lines. The FWHM beam widths are 29 arcsec for the former pair of
lines and 21 arcsec for the latter four.

\begin{figure*}
\includegraphics[width=12cm]{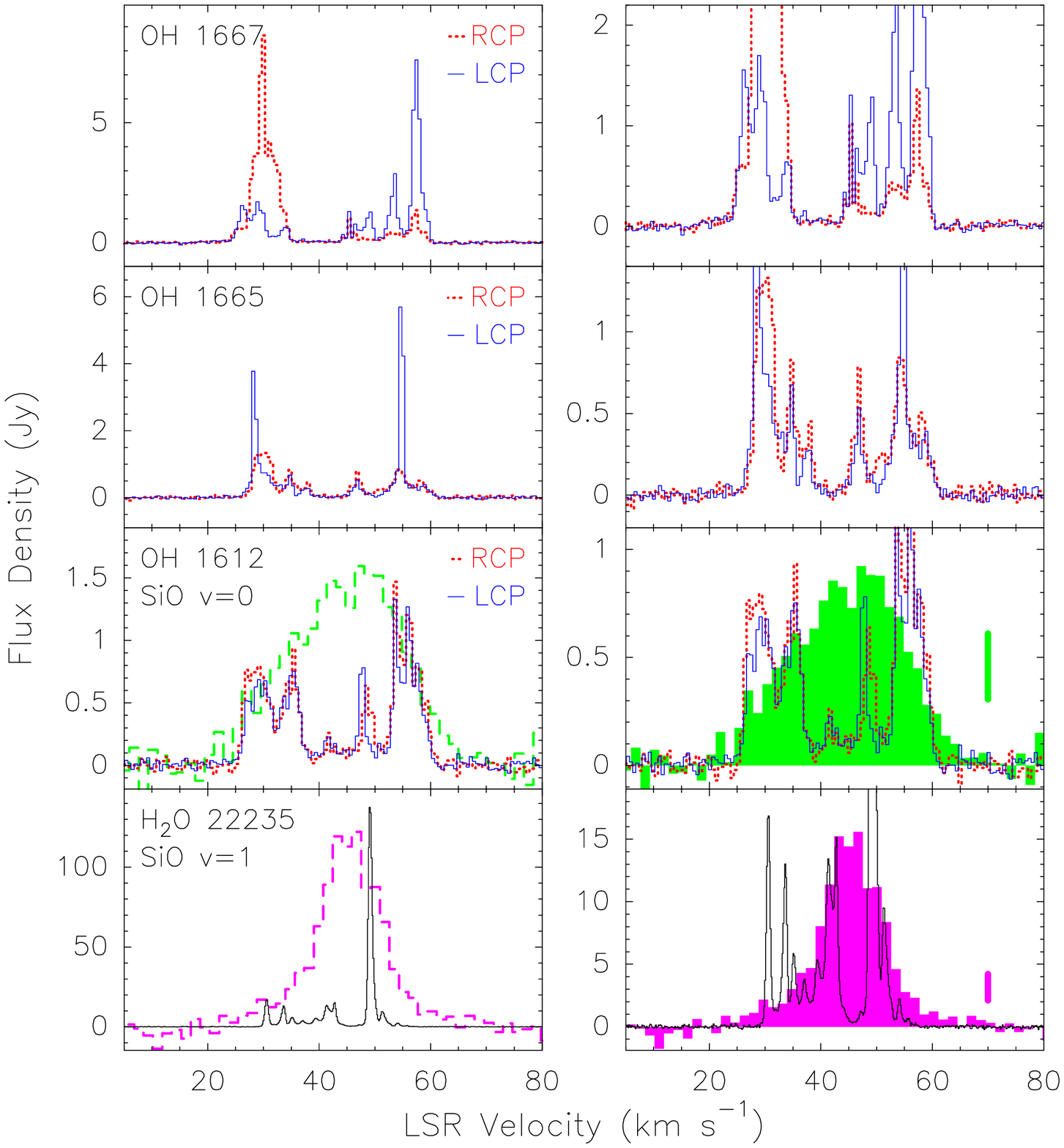}
\centering \caption{Molecular lines observed toward IRC\,$-$10414:
The left column shows, bottom to top, spectra of the 22.2 GHz
H$_2$O and the 1612, 1665 and 1667 MHz OH maser transitions taken
with the Effelsberg 100 m telescope. For the OH lines, the LCP and
RCP spectra shown as full and dotted lines, respectively. The
H$_2$O and 1612 MHz OH spectra are overlaid on spectra of the SiO
217.1 GHz $v=0, J=5-4$ and 215.6 GHz $v=1, J=5-4$ transitions (in
dashed lines respectively). The right column shows the same with
the strongest H$_2$O and OH maser features truncated to emphasize
lower intensity emission and the SiO spectra represented as shaded
areas. The vertical bars define the flux density scale for the SiO
lines and correspond, in both cases, to 1 Jy.} \label{fig:eff}
\end{figure*}

The centimeter wavelength observations were made with the MPIfR
Effelsberg 100 m radio telescope. The 22.2 GHz observations of the
H$_2$O maser line (made on 2012 July 4) and the 1.7 GHz OH
transitions (2012 June 10) used the facility K- and L-band
receivers, respectively. For both sets of observations, the
facility FFTS afforded two 100 MHz wide modules, each with 32768
frequency channels. One channel detected left, the other right
circularly polarized radiation. For the OH lines, the channel
spacing corresponds to $0.55 \, \kms$ in velocity units, for the
H$_2$O line to $0.04 \, \kms$. For Fig.\,\ref{fig:eff} the
spectrum of the latter was smoothed to $0.16 \, \kms$. To
calibrate the spectra, corrections for the atmospheric attenuation
as well as for the gain-elevation effect were applied. Finally,
the spectra were converted into flux density by applying the
aperture efficiencies for both receivers (L-band: 0.53, K-band:
0.32). The aperture efficiencies were checked by observations of
known flux density calibrators like 3C\,286, 3C\,48 and NGC\,7027
(see Baars et al. 1977).

\begin{table*}
\begin{minipage}{0.7\textwidth}
\caption{Results of molecular line observations.} \label{tab:apex}
\begin{tabular}{@{}lcrrcccccccccccccc@{}}
 \hline
Species & Transition & Frequency & $E_l$&Polarization & $v$-range & $\int S {\rm d}v$ \\
&  & (MHz) &  (K)& & ($\kms$) & (Jy $\kms$)\\ \hline
SiO             & $v=0, J=5-4$ & 217104.98 & 20.8 & & [22,69] & 72.1(1.1) \\
                    & $v=1, J=5-4$ & 215595.95 & 1790.0 & & [22,68] & 104.6(1.7) \\
                    & $v=0, J=7-6$ & 303926.81 & 43.8 & & $\approx$[20,68] & 103.0(1.2) \\
                    & $v=1, J=7-6$ & 301814.33 & 1812.8 & & $\sim$[21,61] & 113.2(3.0) \\
SO      &$J_K = 7_7-6_6$ & 301286.12 & 56.5 & & $\sim$[24,63] & 12.8(1.4) \\
                   &$J_K = 7_8-6_7$ & 304077.84 & 47.6 & & $\sim$[28,62] & 14.3(1.5) \\
H$_2$O  & $6_{16} - 5_{23} $ & 22235.08 & 642.5 & LCP+RCP & [29,58] & 226(1) \\
OH              & $F=1-2$& 1612.231     & 0.0& LCP & [23,62] & 13.7(0.2) \\
                    &           &           &   & RCP & [25,63] & 15.2(0.2) \\
                & $F=1-1$& 1665.408     & 0.0   & LCP & [23,63] & 17.5(0.2) \\
                    &           &                   &   & RCP & [26,62] & 13.5(0.2) \\
                    & $F=2-2$& 1667.359     & 0.0& LCP & [20,61] & 32.0(0.2) \\
                    &     &                         &   &RCP & [22,62] & 34.4(0.2) \\
                    & $F=2-1$& 1720.530     & 0.0&LCP & [20,63] & $<0.9$ \\
                    &     &                         &   &RCP & [20,63] & $<0.9$ \\
\hline
\end{tabular}\\
Columns are: molecule, quantum number of transition, rest
frequency, energy above ground state of lower energy level, (for
OH:) sense of polarization, velocity range covered by line (FWZP)
and integrated flux density. For the H$_2$O line the quantum
numbers are $J_{K_{a}K_{c}}$. The OH lines are magnetic hfs
transitions between sublevels of the $^2\Pi_{3/2}, J=3/2$
rotational ground state. For them, flux densities in left and
right circular polarization (LCP and RCP) are listed. For the
non-detected OH 1720 MHz line the quoted flux density upper limit
has been calculated from three times the rms noise value of the
spectrum and assuming a line width equal to that of the detected
OH transitions, i.e. [20,63] $\kms$.\\
\end{minipage}
\end{table*}

\subsection{Molecular line results and discussion}
\label{sec:mas-res}

The results of our spectral line observations are presented in
Table\,\ref{tab:apex}. Fig.\,\ref{fig:sio} shows the spectra of
the SO $J_K = 7_7-6_6$ and $7_8-6_7$ and the SiO $J=5-4$ and
$J=7-6$ transitions from within the $v=0$ and the $v=1$ states
observed with the APEX telescope, while Fig.\,\ref{fig:eff}
presents the Effelsberg spectra of the 22.2 GHz H$_2$O maser
lines, the 1612 MHz OH ``satellite'' hyperfine structure (hfs)
line and the 1665 and 1665 MHz ``main'' hfs lines. For all the OH
lines, both right and left circularly polarized (RCP and LCP)
spectra are shown. The 1720 MHz satellite line was, as is expected
for evolved stars, not detected. From the observed total velocity
spread and its centroid, we determine for IRC\,$-$10414 $v_\infty
=21\pm2 \, \kms$ and $v_{\rm LSR} =43\pm2 \, \kms$ (which
corresponds to a heliocentric radial velocity of $v_{\rm
r,hel}=28.6\pm2.0 \, \kms$). Our value for $v_{\rm LSR}$ is
consistent with that determined by Ukita \& Goldsmith (1984) for
the $v=1, J=2-1$ SiO maser line, while $v_\infty$ is larger than
the figure of $15-16 \, \kms$ inferred by Blommaert et al. (1994)
and Sevenster et al. (2001) from the separation of the red- and
blue-shifted peaks of the OH 1612 MHz line. The reason for the
difference may be that during the epochs at which these authors'
data were taken maser features at velocities closer to $v_{\rm
LSR}$ than $v_\infty$ may have dominated the spectrum.

Notably, the intensity of the OH 1612 MHz line is between 1 and 2
Jy, which is similar to values found during previous observations
(Blommaert et al. 1994; Sevenster et al. 2001). The H$_2$O maser's
flux density exceeds 100 Jy, as during all its previous
measurements (see Kleinman et al. 1978; Lada et al. 1981; Maeda et
al. 2008).

To put these findings in a context, we first briefly summarize
maser emission from evolved, oxygen-rich stars (see Habing 1996
for an extensive treatment). We start with SiO masers, which are
formed within a few stellar radii from the photosphere where the
high density and SiO abundance and an intense infrared radiation
field fulfil the pumping requirements, which require densities
between $10^9$ and $10^{10}$ cm$^{-3}$ and temperatures $> 1000$~K
(Lockett \&\ Elitzur 1992; Bujarrabal 1994), and allow build up of
sufficient maser gain. Further away from the star, SiO depletes
into the forming dust grains although enough remains in the gas
phase to produce thermally excited SiO emission arising from large
parts of the circumstellar envelope and reaching terminal
velocity. H$_2$O maser emission arises from a still relatively
warm region outside of the dust formation region in which the
outflow has reached much or most of $v_\infty$, roughly from the
same region as main-line (1665 and 1667 MHz) OH masers. In
contrast, the satellite (1612 MHz) OH masers arise from a much
larger region. The above picture, well illustrated by fig.\,1 of
Chapman \& Cohen (1986) for the case of the RSG VX\,Sgr, is
consistent with observations of the few known OH emitting RSGs
(i.e., VX\,Sgr, VY\,CMa, NML\,Cyg) and so-called OH/IR stars,
which are  typically AGB stars of more than one solar mass with
higher mass-loss rates than Mira variables and, consequently,
optically opaque envelopes. Solar mass oxygen-rich AGB stars
(``Miras''), which have lower mass-loss rates than RSGs, usually
don't show 1612 MHz emission at all. All maser lines show
significant variability. In particular, for Miras the narrow
features in SiO maser spectra completely change from one stellar
cycle (of $\sim 300$ d or less) to the next.

The maser phenomenology of  IRC\,$-$10414 shows clear deviations
from the canonical picture outlined above. SiO masers from both
RSGs and Miras from the $v=1$ or higher vibrational state
typically show several narrow (from $\sim 1$ to a few $\kms$ wide)
maser features that group around $v_{\rm LSR}$ within a velocity
interval that is significantly smaller than $2v_\infty$. SiO lines
from the vibrational ground state ($v=0$) mostly have smooth,
parabolic profiles, indicating spatially unresolved optically
thick emission, in some cases (for the $v=0, J=1-0$ line)
superposed with a few maser spikes. In the case of IRC\,$-$10414,
both the $v=0, J=5-4$ and $7-6$ lines behave as expected,
showing(almost) parabolic profiles, with a large full width at
zero power (FWZP) of $\approx2v_\infty$. Remarkably, the $v=1$
lines have similar widths, but more asymmetric, smooth (not spiky)
profiles that are more strongly skewed to higher velocity than
those of the $v=0$ lines. While the profile and intensity of the
$v=0$ lines are consistent with thermal emission with a small beam
filling factor (given its 0.1 K main-beam brightness temperature),
the $v=1$ line {\it must definitely} be masing and arise from an
even more compact region. This is clear from the densities and
temperatures required for its excitation (see above). With the
caveat that their spectra were taken 30 yr apart, we note that the
$v=1, J=2-1$ line observed by Ukita \& Goldsmith (1984) in 1982
April was roughly twice as strong as the $v=1, J=5-4$ one when we
observed it. This is in line with the findings of Jewell et al.
(1987) who find for a diverse sample of 18 Miras and RSGs flux
ratios between 1 and 50. We note that, in contrast to
IRC\,$-$10414, all the spectra shown for the $v=1, J=2-1$ and
$J=5-4$ lines by Jewell et al. are dominated by one or a few
narrow features. For the $v=1, J=2-1$ line toward IRC\,$-$10414,
Ukita \& Goldsmith (1984) observe a similarly smooth (but noisier)
shape we do for the $v=1, J=5-4$ line, while a more complex
profile for the $v=1, J=2-1$ line (smooth with a narrower peak)
was observed by Imai, Deguchi \& Miyoshi (1999), who also present
spectra of the $v=1$ and 2, $J=1-0$ lines (the latter being
uncharacteristically weak).

Remarkably, the H$_2$O maser line and the OH main (1665 and 1667
MHz) and satellite (1612 MHz) hfs lines all cover a similar
velocity range, i.e, $2v_\infty$, with a large number of narrow
maser features. All three of the OH lines show narrow emission
features over the  velocity range and enhanced emission at high
and low velocities, close to $\pm v_\infty$. In contrast, the vast
majority of 1612 MHz spectra of typical OH/IR stars are
characterised by two strong spikes at $\pm v_\infty$ with a steady
drop off toward the center of the double-peaked profile and an
emission minimum at $v_{\rm LSR}$. Almost all of the 286 OH/IR
stars detected in their VLA survey, for which Sevenster et al.
(2001) show spectra, have this appearance. Although the low
signal-to-noise ratio is an issue for some sources, by visual
inspection of their fig.\,10 we only find about 10 sources with
more than two features (and 10 are single-featured). In contract,
our much higher quality 1612 MHz spectrum of IRC\,$-$10414 rivals
in complexity that of the ``classical'' OH/IR red super (or even
hyper) giants VY\,CMa, VX\,Sgr and IRC\,+10420 (see Cohen et al.
1987 for high quality/high resolution spectra of these sources).
These sources have $v_\infty$ of 25 to $50 \, \kms$. The terminal
wind velocity of IRC\,$-$10414 of $21 \, \kms$ is somewhat smaller
than these values, but larger than the median $v_\infty$ of $15 \,
\kms$ determined by Sevenster et al. (2001) for their sample of
OH/IR stars.

In the bands covering the SiO $J=7-6$ lines, we also detect weak
emission from two SO lines. Since their level energies above the
ground state are similar and because of the limited
signal-to-noise ratios of their spectra, it is not possible to use
these lines' ratio for calculating a rotation temperature. SO is a
well known chemical constituent of the circumstellar envelopes of
RSGs and O-rich AGB stars (Sahai \&\ Wannier 1992; Omont et al.
1993; Tenebaum et al. 2010).

In summary, the complexity and width of the maser spectra observed
for IRC\,$-$10414 support the RSG nature of the star.

\section{Discussion}
\label{sec:dis}

\subsection{IRC\,$-$10414}
\label{sec:star}

\subsubsection{Parameters of IRC\,$-$10414} \label{sec:par}

Interpolation of the new effective temperature, $T_{\rm eff}$,
scale of Galactic RSGs by Levesque et al. (2005) to M7 supergiants
gives $T_{\rm eff} \approx3300$ K for IRC\,$-$10414. This figure
agrees well with that derived from the relation between $T_{\rm
eff}$ and the spectral type of G$-$M-type supergiants (van Belle,
Creech-Eakman \& Hart 2009):
\begin{equation}
T_{\rm eff}=-(123\pm25)SpT+(4724\pm175) \, {\rm K},
\label{eqn:tem}
\end{equation}
were $SpT$=6$...$14 correspond to M0$...$M8, which for M7 stars
($SpT=13$) gives $T_{\rm eff} =3125\pm369$ K.

With $T_{\rm eff} =3300$ K, one can calculate the intrinsic
$(J-K)_0$ colour of IRC\,$-$10414 and its bolometric $K$-band
correction, BC$_K$, using the relations (Levesque et al. 2005):
\begin{equation}
(J-K)_0 =3.10-0.547(T_{\rm eff}/1000 \, {\rm K}) \label{eqn:j-k}
\end{equation}
and
\begin{equation}
{\rm BC}_K=5.574 - 0.7589(T_{\rm eff}/1000 \, {\rm K}) \, ,
\label{eqn:bc}
\end{equation}
which give $(J-K)_0 =1.29$ mag and BC$_K$=3.07 mag. Then, using
equations\,(\ref{eqn:j-k}) and (\ref{eqn:bc}), and the $J$ and
$K_{\rm s}$ magnitudes from Table\,\ref{tab:det}, one can estimate
the $K$-band extinction towards the star, its $K$-band absolute
magnitude, and the bolometric luminosity:
\begin{equation}
A_K =0.66[(J-K)-(J-K)_0] \, , \label{eqn:ak}
\end{equation}
\begin{equation}
M_K =K-DM-A_K \, , \label{eqn:mk}
\end{equation}
\begin{equation}
\log(L/{\rm L}_\odot)=0.4(4.74-M_K-{\rm BC}_K) \, ,
\label{eqn:lum}
\end{equation}
where $K$=$K_{\rm s}$+$0.04$ mag (Carpenter 2001) and $DM$ is the
distance modulus ($DM\approx11.51$ mag for $d=2$ kpc). For
$J=2.845$ mag and $K_{\rm s}=0.713$ mag, one has $A_K \approx
0.53$ mag (which corresponds to the visual extinction of $A_V
\approx5$ mag; Rieke \& Lebofsky 1985), $M_K \approx -11.3$ mag
and $\log(L/\lsun)\approx 5.2$. Depending on the initial rotation
velocity of IRC\,$-$10414, the derived luminosity corresponds to
an initial (zero-age main-sequence) mass of the star of
$M\sim20-25 \, \msun$ and implies an age of $\sim 6-10$ Myr
(Ekstr\"{o}m et al. 2012).

$M_K$ can also be estimated from the period-$M_K$ relations of
RSGs proposed by Kiss, Szab\'{o} \& Bedding (2006) and Yang \&
Jiang (2012). Using these relations and assuming that $P=768.16$ d
is the fundamental mode of radial pulsation\footnote{Note that the
long secondary period of 2726.43 d (detected in the light curve of
IRC\,$-$10414 by Richards et al. 2012) might correspond to the
convective turnover time of giant convection cells in the stellar
envelope (Stothers \& Leung 1971; Kiss et al. 2006).}, one finds
the mean value of $M_K\approx-11.4$ mag. One can also estimate the
luminosity of IRC\,$-$10414 using empirical period-luminosity
relation for semi-regular supergiant variables by Turner et al.
(2006; see their fig.\,9), which yields $\log(L/{\rm
L}_\odot)\approx 5.2$. The good agreement of these estimates with
those derived from equations\,(\ref{eqn:ak})--(\ref{eqn:lum})
provides support for our choice of $d=2$ kpc.

Then, using the luminosity-radius-temperature relation by Levesque
et al. (2007):
\begin{eqnarray}
R/R_\odot =(L/\lsun)^{0.5} (T_{\rm eff}/5770 \, {\rm K})^{-2} \, ,
\nonumber
\end{eqnarray}
one finds the radius of IRC\,$-$10414 of $R\approx 1200$
R$_{\odot}$, which makes it one of the biggest known stars (cf.
Wing 2009). For this radius and $M=20-25 \, \msun$, one finds
$\log g\approx -(0.3\div0.4)$ (cf. Section\,\ref{sec:spec-irc}).
It is also instructive to use these $M$ and $R$ and the
period-radius-mass relation for pulsating stars by Gough, Ostriker
\& Stobie (1965),
\begin{eqnarray}
P\approx 10^{-2} \, {\rm d} \,(R/R_{\odot})^2 (M/\msun)^{-1} \, ,
\nonumber
\end{eqnarray}
to derive the expected period of pulsations of IRC\,$-$10414,
$P\approx600-700$ d, which agrees with our choice of $P=768.16$ d
as the fundamental mode of radial pulsations of IRC\,$-$10414.

Recently, Davies et al. (2013) showed that $T_{\rm eff}$ of RSGs
derived from optical and near-infrared spectral energy
distributions might be significantly (by several hundreds of K)
warmer than those based on the $T_{\rm eff}$ scale by Levesque et
al. (2005). Adopting $T_{\rm eff} =3700$ K for IRC\,$-$10414 and
using equations\,(\ref{eqn:j-k})$-$(\ref{eqn:lum}), one finds
$M_{\rm bol}=-8.66$ mag and $\log(L/\lsun)=5.4$. This higher
luminosity, however, would be difficult to reconcile with the
moderate $\dot{M}$ derived for IRC\,$-$10414 in
Section\,\ref{sec:mas}.

\subsubsection{IRC\,$-$10414 as a runaway}
\label{sec:run}

\begin{table*}
\caption{Proper motion, heliocentric radial velocity, peculiar
transverse (in Galactic coordinates) and radial velocities, and
the total space velocity of IRC\,$-$10414.} \label{tab:prop}
\centering
\begin{tabular}{ccccccc}
\hline $\mu _\alpha \cos \delta$ & $\mu _\delta$ & $v_{\rm r,hel}$
& $v_{\rm l}$ & $v_{\rm b}$ & $v_{\rm r}$ & $v_\ast$ \\
(mas ${\rm yr}^{-1}$) & (mas ${\rm yr}^{-1}$) & ($\kms$) & ($\kms$) & ($\kms$) & ($\kms$) & ($\kms$) \\
\hline
$5.4\pm2.9$ & $1.6\pm2.4$ & $28.6\pm2.0$ & $53.9\pm23.9$ & $-30.8\pm26.5$ & $20.6\pm2.0$ & $65.4\pm23.3$ \\
\end{tabular}
\end{table*}

The bow shock interpretation of the arc-like nebula implies that
IRC\,$-$10414 is moving supersonically with respect to the local
ISM. To estimate the peculiar (space) velocity of IRC\,$-$10414
and thereby to check its runaway status, we searched for proper
motion measurements for this star using the VizieR catalogue
access tool\footnote{http://webviz.u-strasbg.fr/viz-bin/VizieR}.
We found several measurements of which the most recent one (and
the one with the smallest claimed errors) is provided by the
fourth U.S. Naval Observatory CCD Astrograph Catalog (UCAC4;
Zacharias et al. 2013). This measurement is given in
Table\,\ref{tab:prop} along with the heliocentric radial velocity
(as measured in Section\,\ref{sec:mas-res}), the components of the
peculiar transverse velocity (in Galactic coordinates), $v_{\rm
l}$ and $v_{\rm b}$, the peculiar radial velocity, $v_{\rm r}$,
and the total space velocity, $v_\ast$, of the star. To derive
these velocities, we used the Galactic constants $R_0 = 8.0$ kpc
and $\Theta _0 =240 \, \kms$ (Reid et al. 2009) and the solar
peculiar motion $(U_{\odot},V_{\odot},W_{\odot})=(11.1,12.2,7.3)
\, \kms$ (Sch\"onrich, Binney \& Dehnen 2010). For the error
calculation, only the errors of the proper motion and the radial
velocity measurements were considered. The obtained space velocity
of $\approx 70\pm20 \, \kms$ implies that IRC\,$-$10414 is a
classical runaway star (e.g. Blaauw 1961).

\begin{figure}
\includegraphics[width=8cm]{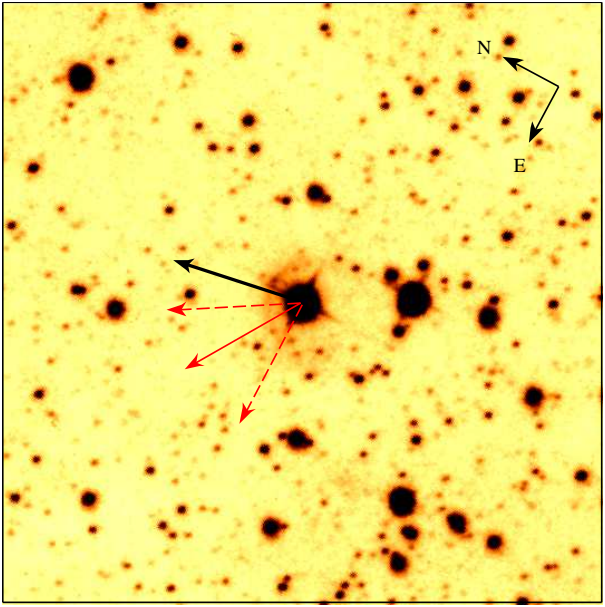}
\centering \caption{SHS image of IRC\,$-$10414 and its bow shock.
The thin (red) arrow shows the direction of motion of the star as
suggested by the proper motion measurement (with 1 sigma
uncertainties shown by dashed arrows). The thick (black) arrow
shows the symmetry axis of the bow shock. See text for details.}
\label{fig:vel}
\end{figure}

Fig.\,\ref{fig:vel} shows that the vector of the peculiar
(transverse) velocity is misaligned with the symmetry axis of the
bow shock. This misalignment might be caused by inaccuracy of the
proper motion measurement or by the effect of a regular flow in
the local ISM. The latter explanation is less likely because it
requires the presence of gas flowing with a velocity of several
tens of $\kms$. Such high-velocity flows could be found in the
close vicinity of star clusters (e.g. Bally et al. 2006; Povich et
al. 2008; Kobulnicky, Gilbert \& Kiminki 2010), but they should be
rare in the field. The quite large margin of error in velocity
components (see Table\,\ref{tab:prop}) leaves the possibility that
IRC\,$-$10414 is moving almost parallel to the Galactic plane (in
the direction of growing Galactic longitude), which is suggested
by the orientation of the symmetry axis of the bow shock as well.
From Table\,\ref{tab:prop} it follows that IRC\,$-$10414 is moving
away of us and that the vector of its space velocity makes an
angle of $\approx 18_{-6} ^{+13}$ degrees with respect to the
plane of the sky. Thus, the inclination of the bow shock does not
much affect our estimate of $R_{\rm SO}$ given in
Section\,\ref{sec:arc} (see Gvaramadze et al. 2011c for details).

\begin{figure}
\includegraphics[width=8cm]{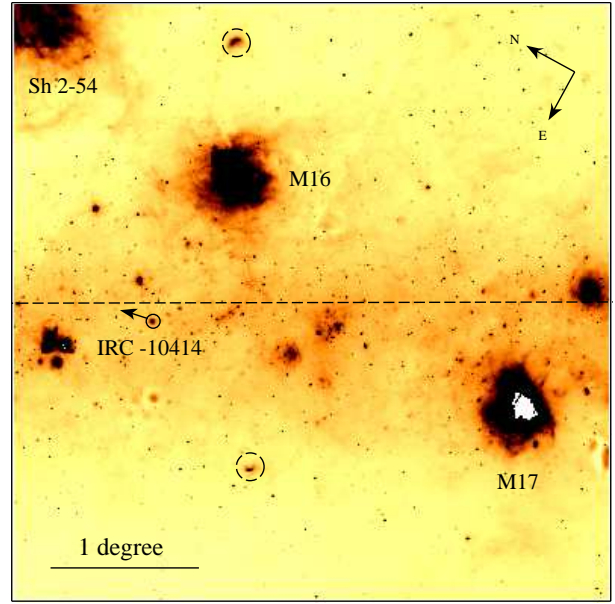}
\centering \caption{$4\degr \times 4\degr$ {\it WISE} 22 $\mu$m
image of the field containing IRC\,$-$10414 and three young
massive star clusters NGC\,6604, NGC\,6611 and NGC\,6618, embedded
in the \hii regions Sh\,2-54, M16 and M17, respectively (the
latter \hii region is highly saturated). The position of
IRC\,$-$10414 is marked by a circle. The arrow shows the direction
of motion of IRC\,$-$10414, as suggested by the symmetry axis of
the bow shock (see Fig.\,\ref{fig:vel}). Two bow shocks produced
by O stars running away from NGC\,6611 (Gvaramadze \& Bomans 2008)
are indicated by dashed circles. The Galactic plane is shown by a
dashed line. See text for details. At a distance of 2 kpc, 1
degree corresponds to $\approx$34.4 pc.} \label{fig:clu}
\end{figure}

Fig.\,\ref{fig:clu} shows the $4\degr \times 4\degr$ {\it WISE}
22\,$\mu$m image\footnote{The image was generated by the NASA's
SkyView facility (McGlynn, Scollick \& White 1996).} of the
Galactic plane centred at $l=16\fdg5, b=0\degr$, with three
prominent \hii regions, Sh\,2-54, M16 and M17, and their central
star clusters, NGC\,6604, NGC\,6611 and NGC\,6618, respectively.
All three clusters are located at about the same distance of
$\approx 2$ kpc (Reipurth 2008; Hillenbrand et al. 1993; Xu et al.
2011) and are believed to originate from the same giant molecular
cloud (Sofue et al. 1986). In Fig.\,\ref{fig:clu} we indicate by
dashed circles the positions of two bow-shock-producing O stars.
Proper motion measurements for these stars and the orientations of
their bow shocks are consistent with the possibility that both
stars are running away from NGC\,6611 (Gvaramadze \& Bomans 2008).
Although the past trajectory of IRC\,$-$10414 (inferred from the
orientation of the peculiar transverse velocity of this star)
might intersect the trajectory of NGC\,6611 as well (see fig.\,1
in Gvaramadze \& Bomans 2008), we note that the age of the cluster
is a factor of $\approx2-3$ younger than that of the star
(Gvaramadze \& Bomans 2008 and references therein), which implies
that NGC\,6611 cannot be the birth place of IRC\,$-$10414.
Instead, assuming that the symmetry axis of the bow shock better
reflects the direction of motion of the star, one can infer that
IRC\,$-$10414 is moving away from M17.

It is, however, unlikely that M17 is the birth place of
IRC\,$-$10414 because the central cluster of this \hii region,
NGC\,6618, is very young ($\sim 1$ Myr; Chini \& Hoffmeister
2008). One can therefore suggest that IRC\,$-$10414 has been
ejected from the more distant stellar system, e.g. from the
Sgr\,OB4 association (Humphreys 1978), which is located further
southwest of M17 (at $\approx 5\fdg5$ or $\approx 190$ pc in
projection from IRC\,$-$10414) and whose distance from the Sun is
also $\approx 2$ kpc (e.g. Mel'nik \& Dambis 2009). In this case,
the ejection event must have happened $\approx3.5_{-1.1} ^{+2.8}$
Myr ago, provided that the peculiar transverse velocity of
IRC\,$-$10414 is $\approx v_l$ (see Table\,\ref{tab:prop}). The
birth place of IRC\,$-$10414 might even be more distant
($\approx9-15$ degrees) if this 6$-$10 Myr old star was ejected
dynamically at the very beginning of dynamical evolution of its
parent stellar system.

Finally, one cannot exclude the possibility that IRC\,$-$10414 was
a member of a runaway binary or hierarchical triple system, which
was dissolved either because of supernova explosion of one of the
binary components (Pflamm-Altenburg \& Kroupa 2010) or evolution
of the most massive members of the triple (Gvaramadze \& Menten
2012). In both cases, IRC\,$-$10414 cannot be traced back to its
parent star cluster (cf. Gvaramadze et al. 2012), while in the
second one, WR\,114 might be the former member of the dissolved
triple system\footnote{Note that the VIIth Catalogue of Galactic
Wolf-Rayet Stars by van der Hucht (2001) gives for WR\,114 a
distance of 2 kpc and indicates this star as a possible binary
system (see, however, Sander, Hamann \& Todt 2012).}. Dissolution
of the triple system would cause IRC\,$-$10414 and WR\,114 to move
in the opposite directions to each other, but this cannot be
proved at present because the existing proper motion measurements
for WR\,114 are highly unreliable.

\subsubsection{Mass-loss rate of IRC\,$-$10414} \label{sec:mas}

Now we use $v_*$ derived in Section\,\ref{sec:run} and the upper
limit on $n_{\rm e}$ obtained in Section\,\ref{sec:bow} to
estimate $\dot{M}$ of IRC\,$-$10414.

First, we use $n_{\rm e}$ to constrain the preshock number
density, $n_0$, i.e. the number density of the local ISM. For this
we use an empirical relation between the [S\,{\sc ii}] electron
number density and the shock velocity $v_{\rm sh}$ (Dopita 1978):
\begin{equation}
n_0 \approx 0.02n_{\rm e} \left({v_{\rm sh} \over 100 \,
\kms}\right)^{-2} \, .
\label{eqn:dop}
\end{equation}
With $n_{\rm e} \leq 100 \, {\rm cm}^{-3}$ and $v_{\rm sh}=v_\ast
=70 \, \kms$, one has $n_0 \leq 5 \, {\rm cm}^{-3}$.

Then, we express $\dot{M}$ through $n_0$ and the observables,
$v_\ast$, $R_{\rm SO}$ and $v_\infty$:
\begin{eqnarray}
\dot{M} _{\rm obs} = 1.0\times 10^{-5} \myr \left({{\it v}_\ast
\over 70 \, \kms}\right)^2 \left({{\it R}_{\rm SO} \over 0.14 \, {\rm pc}}\right)^2 \nonumber \\
\times \left({n_0 \over 5 \, {\rm cm}^{-3}}\right)\left({v_\infty
\over 21 \, \kms}\right)^{-1} \, . \nonumber \label{eqn:mass-loss}
\end{eqnarray}
This estimate of $\dot{M}$ should be considered as
an upper limit because $n_0$ is constrained only from the upper
end. Note also that, according to equation\,(\ref{eqn:dop}), $n_0
\propto v_\ast ^{-2}$, so that formally $\dot{M} _{\rm obs}$ does
not depend on the stellar space velocity. In fact, the dependence
of $n_0$ on $v_\ast$ is more complex because $n_{\rm e}$ is a
function of $v_\ast$ (or $v_{\rm sh}$) as well.

It is interesting to compare $\dot{M} _{\rm obs}$ with rates based
on various mass-loss prescriptions proposed for RSGs (see Mauron
\& Josselin 2011 for a recent review). Using the recipe by Jura \&
Kleinmann (1989) and the parameters of IRC\,$-$10414, one finds
$\dot{M}\approx 2\times 10^{-5} \myr$ (cf.
Section\,\ref{sec:irc}), which is at least a factor of 2 higher
than $\dot{M} _{\rm obs}$.

A lower mass-loss rate follows from the use of the empirical law
by de Jager, Nieuwenhuijzen \& van der Hucht (1988). For
$L=10^{5.2} \, \lsun$ and $T_{\rm eff} =3300$ K, the de Jager et
al.'s prescription predicts $\dot{M}\approx 5\times10^{-6} \,
\myr$ (see fig.\,6 in Mauron \& Josselin 2011). This estimate
would be consistent with $\dot{M} _{\rm obs}$ if $n_0 =2.5 \, {\rm
cm}^{-3}$.

An even lower rate could be derived from the recipe by Verhoelst
et al. (2009):
\begin{eqnarray}
\dot{M}=10^{-16.2\pm1.8} (L/\lsun)^{1.89\pm0.36} \, , \nonumber
\end{eqnarray}
which gives $\dot{M}=4.2\times10^{-7} \, \myr$. Although this
estimate would agree with $\dot{M} _{\rm obs}$ for a reasonable
number density of the local ISM of $n_0 =0.2 \, {\rm cm}^{-3}$, we
note that the large spread allowed by the Verhoelst et al.'s
recipe leaves the possibility that the actual $\dot{M}$ could be
much higher.

$\dot{M}$ of IRC\,$-$10414 can also be estimated through the
$K_{\rm s}-$[24] colour of this star, where [24] is the MIPS
24\,$\mu$m magnitude (see Bonanos et al. 2010 and references
therein). The saturation of the MIPS 24\,$\mu$m image of
IRC\,$-$10414 (see Fig.\,\ref{fig:arc}), however, does not allow
us to measure [24]. Hence, we use the {\it IRAS} 25\,$\mu$m
magnitude (Helou \& Walker 1988), [25]=$-4.213$ mag, instead of
[24]. Assuming the gas-to-dust ratio of 200 (typical of Galactic
RSGs) and using equations\,(1) and (2) in Bonanos et al. (2010),
one finds $\dot{M}=9.9\times10^{-6} \myr$, which agrees well with
$\dot{M} _{\rm obs}$ (provided that $n_0\approx5 \, {\rm
cm}^{-3}$).

Finally, $\dot{M}$ can be derived from the prescription by van
Loon et al. (2005), according to which $\dot{M}$ increases almost
linearly with increasing $L$ and sharply decreases with increase
of $T_{\rm eff}$:
\begin{eqnarray}
\dot{M}=2.5\times 10^{-5} \myr (L/10^5 \, \lsun)^{1.05} (T_{\rm
eff}/3500 \, {\rm K})^{-6.3} \, . \nonumber
\label{eqn:ma-lo}
\end{eqnarray}
For $L=10^{5.2} \, \lsun$ and $T_{\rm eff}=3300$ K this
formulation gives $\dot{M}=5.9\times10^{-5} \myr$, which is much
higher than $\dot{M} _{\rm obs}$ (recall that $\dot{M} _{\rm obs}$
is the upper limit on the mass-loss rate of IRC\,$-$10414).
Interestingly, the mass-loss rates of both known
bow-shock-producing RSGs, Betelgeuse and $\mu$\,Cep, are much
smaller than those predicted by van Loon et al. (2005) as well.

\subsection{Bow shock}
\label{sec:bow}

\subsubsection{Stability}
\label{sec:stab}

Numerical and analytical studies of bow shocks produced by RSG and
other cool stars show that they are unstable to a considerable
degree (e.g. Brighenti \& D'Ercole 1995; Dgani, van Buren \&
Noriega-Crespo 1996; Wareing, Zijlstra \& O'Brien 2007a,b; van
Marle et al. 2011; Mohamed, Mackey \& Langer 2012; Cox et al.
2012; Decin et al. 2012). Correspondingly, one can expect that
these bow shocks should have a ragged appearance, which is in an
obvious conflict with the observed smoothness of bow shocks
associated with Betelgeuse and IRC\,$-$10414.

A possible solution proposed by Mohamed et al. (2012) is that
younger bow shocks are more stable (because all instabilities have
a finite growth timescale), so a very smooth bow shock can be an
indication of its youth. If the bow shock is young then it is
likely moving into gas from previous mass-loss phases, so the
circumstellar gas could already be enriched in elements such as
nitrogen. Mackey et al. (2012) showed that a BSG evolving to a RSG
would have a low-mass bow shock expanding into ionized BSG wind
material for a short time ($\approx 20\,000$ years) after the star
evolves to a RSG. It is possible but unlikely that we observe
IRC\,$-$10414 in just this phase of evolution.

It should be noted that in the above-mentioned papers, the stellar
wind was considered to be neutral (which is natural for cool, RSG
and AGB, stars), while the ISM was assumed to be either neutral or
ionized. Also, numerical modelling of bow shocks generated by hot
stars in the ionized ISM show that they are stable for certain
$v_*$ and $\dot{M}$ (e.g. Comer\'{o}n \& Kaper 1998; Meyer et al.,
in preparation). Thus, one can suppose that the smooth shape of
the bow shocks generated by IRC\,$-$10414 and Betelgeuse is
because the wind of these stars and their ambient ISM are both
ionized. This supposition can be supported by the following
considerations.

Mohamed et al. (2012) found that their simulated bow shocks were
much less stable than those of Wareing et al. (2007a) for similar
wind and ISM parameters. The only physical difference in the
models was that Wareing et al. (2007) assumed ``an unphysical
temperature of $10^4$ K" for the stellar wind, whereas Mohamed et
al. (2012) allowed gas to cool to lower temperatures. Mohamed et
al. proposed that the greater compression caused by the cooling in
their simulations made the bow shock thinner and therefore less
stable. Proceeding from this, one can expect that if the wind and
ISM are both photoionized, so that their temperature is kept at
$T\approx10^4$ K and thereby prevented from cooling to lower
temperatures, then the bow shock would be more stable than a bow
shock that is not photoionized. This will be studied in more
detail for the bow shock of IRC\,$-$10414 by Meyer et al. (in
preparation).

While an ionized ISM is the most widespread component of the ISM
(McKee \& Ostriker 1977), the assumption that the wind of
IRC\,$-$10414 is ionized is less trivial because this star is too
cool to ionize the wind material by itself. The wind of
IRC\,$-$10414 could be ionized (at least partially) if this star
forms a binary system with a hot massive star or a compact object.
Such a situation takes place in the case of $\alpha$\,Sco
(Antares), which is composed of M1\,Iab and B3\,V: stars
(Kudritzki \& Reimers 1978; Reimers et al. 2008). In the absence
of data supporting the possible binarity of IRC\,$-$10414, we
consider an alternative possibility, namely that the wind of this
star is ionized by an external source of ionizing photons (see
next section). A good evidence supporting this possibility is the
presence of the very strong [N\,{\sc ii}] emission in the spectrum
of the bow shock (Section\,\ref{sec:spec-bow}), which means that
the stellar wind is ionized to a significant degree. Since the
wind material cannot be collisionally ionized because the reverse
shock is too weak, it is natural to assume that it is photoionized
by some external source(s).

\subsubsection{Ionization source}
\label{sec:ion}

The close angular proximity between IRC\,$-$10414 and WR\,114
prompted us to consider the possibility that the latter star
photoionizes the wind of the former one (recall that the distance
to WR\,114 is 2 kpc).

\begin{figure}
\includegraphics[width=8cm]{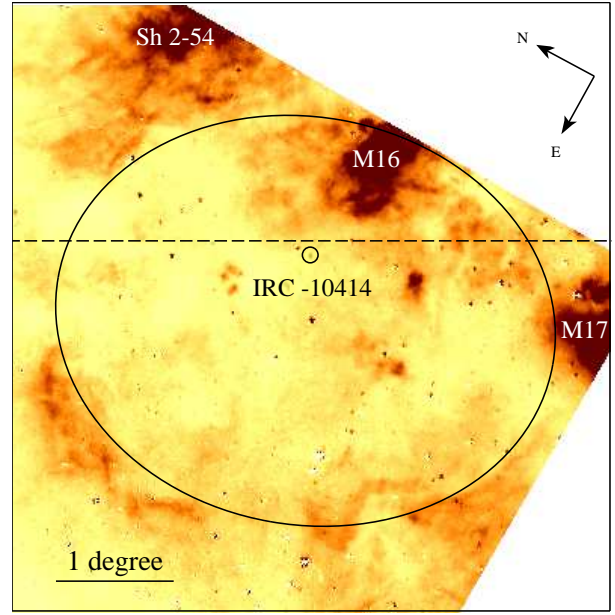}
\centering \caption{SHASSA continuum-subtracted image of the field
containing IRC\,$-$10414 (marked by a circle) and three young
massive star clusters, NGC\,6604, NGC\,6611 and NGC\,6618,
embedded in the \hii regions, Sh\,2-54, M16 and M17, respectively.
The putative ionization-bounded \hii region created by NGC\,6611
is outlined by an ellipse. The Galactic plane is shown by a dashed
line. See text for details. At a distance of 2 kpc, 1 degree
corresponds to $\approx$34.4 pc.} \label{fig:hii}
\end{figure}

Following Morris \& Jura (1983), we can calculate the stand-off
radius of the ionization front, $R_{\rm i}$, in the RSG wind (i.e.
the minimum distance from the ionization front to the RSG star) as
a function of the incident ionizing radiation field:
\begin{eqnarray}
R_{\rm i} = 0.01 \, {\rm pc} \left({\dot{M} \over 2\times10^{-6}
\,
\myr}\right)^{2/3}  \nonumber \\
\times \left({v_\infty\over 21 \,\kms}\right)^{-2/3}
\left({F_\gamma \over 10^{10} \, {\rm cm}^{-2} \, {\rm
s}^{-1}}\right)^{-1/3} \, ,
\label{eqn:ion}
\end{eqnarray}
where $F_\gamma$ is the ionizing photon flux. For a point source
of ionizing photons $F_\gamma =L_\gamma/(4\pi l^2)$, where
$L_\gamma$ is the ionizing photon luminosity of the source and $l$
is the linear separation between the source and the RSG star.

With $L_{\gamma}=10^{48.9} \, {\rm s}^{-1}$ (Crowther 2007) for a
WC5 star like WR\,114, and assuming that $l$ is equal to the
projected distance between the two stars, i.e. $l\approx0.43$ pc,
one finds $F_\gamma\approx 3.6\times10^{11} \, {\rm s}^{-1} \,
{\rm cm}^{-2}$ and $R_{\rm i}\approx3.4\times10^{-3}$ pc or
$\approx0.02R_{\rm SO}$ (i.e. WR\,114 can, in principle, ionize
very deep layers of the wind of IRC\,$-$10414). But such a small
separation between IRC\,$-$10414 and WR\,114 is unlikely because
the strong wind of the later star would prevent formation of the
bow shock around the former one. Thus, the separation between the
two stars should at least be as large as the characteristic scale
of the wind-bubble created by WR\,114, i.e. (Weaver et al. 1977):
\begin{eqnarray}
R_{\rm b} =27 \, {\rm pc} \left({L_{\rm w}\over 10^{36} \, {\rm
erg} \, {\rm s}^{-1}}\right)^{1/5}\left({n_0\over 1 \, {\rm
cm}^{-3}}\right)^{-1/5} \left({t\over 1 \, {\rm Myr}}\right)^{3/5}
\, , \nonumber
\end{eqnarray}
where $L_{\rm w}=\dot{M}v_\infty ^2/2$ is the mechanical
luminosity of the star and $t$ is the age of the bubble. With
$\dot{M}=2.9\times10^{-5} \, \myr$ and $v_\infty =3200 \, \kms$
(Sander et al. 2012) and assuming $t=0.5$ Myr, we obtain $L_{\rm
w}\approx9.4\times10^{37} \, {\rm erg} \, {\rm s}^{-1}$ and
$R_{\rm b}\approx 44$ pc. For $l\geq R_{\rm b}$, one finds $R_{\rm
i} \geq 0.4R_{\rm SO}$, i.e. WR\,114 still could be the ionization
source of the wind of IRC\,$-$10414.

Another possibility is that the wind of IRC\,$-$10414 is ionized
by the star cluster NGC\,6611 (cf. Morris \& Jura 1983), which is
located at $\approx38$ pc in projection from IRC\,$-$10414. Using
the census of massive stars in NGC\,6611 from Evans et al. (2005)
and calibration of stellar parameters of Galactic O stars from
Martins et al. (2005), we estimated the total ionizing photon
luminosity of the cluster to be $L_\gamma({\rm
NGC}\,6611)\approx10^{50} \, {\rm s}^{-1}$ (cf. Hester et al.
1996). Assuming that $l=38$ pc and neglecting attenuation of the
ionizing flux by photoelectric absorption in the ISM, one finds
$F_\gamma\approx 5.8\times10^{8} \, {\rm s}^{-1} \, {\rm
cm}^{-2}$. Then using equation\,(\ref{eqn:ion}), we obtain $R_{\rm
i}\approx0.03$ pc or $\approx0.2R_{\rm SO}$, i.e. NGC\,6611 could
be the ionization source of the wind of IRC\,$-$10414 as well.

The high $L_{\gamma}$ of NGC\,6611 might also be responsible for
creation of an extended \hii region to the east-southeast of
IRC\,$-$10414. The characteristic scale of this \hii region should
be of the same order of magnitude as the Str\"{o}mgren radius,
which is given by (e.g. Lequeux 2005):
\begin{eqnarray}
R_{\rm S} = 140 \, {\rm pc} \left({L_\gamma \over 10^{50}  \, {\rm
s}^{-1}}\right)^{1/3}\left({n_0\over 1 \, {\rm
cm}^{-3}}\right)^{-2/3} \, . \nonumber
\end{eqnarray}
Indeed, inspection of the Southern H-alpha Sky Survey Atlas
(SHASSA; Gaustad et al. 2001) revealed an
$\approx2.7\degr\times3.3\degr$ structure (outlined in
Fig.\,\ref{fig:hii} by an ellipse). The linear size of this
structure of $\approx90\times110$ pc approximately agrees with
$R_{\rm S}$, which suggests that it might be the {\it
ionization-bounded} \hii region produced by NGC\,6611 (as distinct
from the {\it density-bounded} blister-like \hii region M16).

To conclude, we note that if the wind of IRC\,$-$10414 is ionized
by a single ionizing source and if $R_{\rm i}$ constitutes a
non-negligible fraction of $R_{\rm SO}$ (see above), then one can
expect to see a comet-like tail of H\,{\sc i} gas attached to the
star and pointed away from the source of ionization.
Interestingly, observations of Betelgeuse in the H\,{\sc i} line
at 21 cm by Le Bertre et al. (2012) revealed such a tail in the
direction opposite to the star's space motion. This detection may
imply that Betelgeuse is moving towards the ionizing source and
suggests that the stability of the bow shock around this star
could be because a part of the stellar wind is ionized. Similar
observations of IRC\,$-$10414, however, appear to be practically
impossible, because of the much larger distance to this star and
contamination with H\,{\sc ii} emission in the Galactic plane.

\section{Summary}
\label{sec:sum}

In this paper, we presented the discovery of an optically visible
arc-like nebula around the late M star IRC\,$-$10414 using the
SuperCOSMOS H-alpha Survey. We also reported the results of
follow-up spectroscopy of IRC\,$-$10414 and its nebula with the
Southern African Large Telescope (SALT). We have classified
IRC\,$-$10414 as an M7 supergiant, thereby confirming previous
claims on the red supergiant (RSG) nature of this star based on
observations of its maser emission. The RSG nature of
IRC\,$-$10414 was further supported by our new radio- and
(sub)millimeter-wavelength molecular line observations made with
the Atacama Pathfinder Experiment (APEX) 12 meter telescope and
the Effelsberg 100 m radio telescope, respectively. We detected
maser emission from OH, H$_2$O and SiO, as well as thermally
excited emission from the latter and from SO. These observations
also yield the star's local-standard of rest velocity of $43\pm2
\, \kms$ and the wind terminal velocity of $21\pm2 \, \kms$.

Using the recent proper motion measurement for IRC\,$-$10414, we
estimated the space velocity of this star to be $\approx70\pm20 \,
\kms$, which implies that IRC\,$-$10414 is a classical runaway.
This finding along with the arc-like shape of the nebula suggest
that the nebula is a bow shock. The bow shock interpretation was
reinforced by the SALT spectrum of the nebula, which yields the
[S\,{\sc ii}] $\lambda\lambda$6716,6731/H$\alpha$ ratio of 0.4,
typical of shock-excited nebulae. From intensities of the [N\,{\sc
ii}] and [S\,{\sc ii}] lines detected in the spectrum of the bow
shock we found that the line-emitting material is enriched in
nitrogen. We considered this as an indication that the bow shock
emission at least partially originates in the shocked stellar
wind, which in RSGs is overabundant in nitrogen because of the
dredge-up of the nuclear processed gas. Detection of the bow shock
around IRC\,$-$10414 makes this star the third case of
bow-shock-producing RSGs (with two other cases being Betelgeuse
and $\mu$ Cep) and the first one with a bow shock visible in the
optical range.

The spectroscopy of the bow shock also allowed us to constrain the
number density of the local ISM, which along with the stand-off
distance of the bow shock and the wind and space velocities of the
star provide an estimate of the mass-loss rate of IRC\,$-$10414.
This estimate was compared with mass-loss rates based on various
prescriptions proposed for RSGs.

We discussed the smooth appearance of the bow shocks around
IRC\,$-$10414 and Betelgeuse and suggested that one of the
necessary conditions for stability of bow shocks generated by RSG
and other cool stars is the ionization of the stellar wind.
Possible ionization sources of the wind of IRC\,$-$10414 were
proposed and discussed.

\section{Acknowledgements}
AYK acknowledges the support from the National Research Foundation
(NRF) of South Africa. JM is funded by a fellowship from the
Alexander von Humboldt Foundation. We are grateful to E.M.
Levesque for useful discussion. This work was partly based on
observations with the Southern African Large Telescope (SALT) and
the 100-m telescope of the Max-Planck-Institut f\"ur
Radioastronomie at Effelsberg, and supported by the Deutsche
Forschungsgemeinschaft priority program 1573, 'Physics of the
Interstellar Medium'. It also has made use of the NASA/IPAC
Infrared Science Archive, which is operated by the Jet Propulsion
Laboratory, California Institute of Technology, under contract
with the National Aeronautics and Space Administration, the
Southern H-Alpha Sky Survey Atlas (SHASSA), which is supported by
the National Science Foundation, the SIMBAD data base and the
VizieR catalogue access tool, both operated at CDS, Strasbourg,
France.

\end{document}